\documentclass[conference]{IEEEtran}
\IEEEoverridecommandlockouts

\usepackage{cite}
\usepackage{amsmath,amssymb,amsfonts}
\usepackage{algorithmic}
\usepackage{graphicx}
\usepackage{textcomp}
\usepackage{xcolor}
\usepackage{graphicx}
\usepackage{enumitem}
\usepackage{booktabs}
\usepackage{xurl}
\def\BibTeX{{\rm B\kern-.05em{\sc i\kern-.025em b}\kern-.08em
    T\kern-.1667em\lower.7ex\hbox{E}\kern-.125emX}}
\begin{document}

\title{Creating Opportunities: Co-Designing an mHealth App with Older Adults\\
}
\author{
    \IEEEauthorblockN{Abhinav Choudhry\IEEEauthorrefmark{1},
    Bashab Mazumder\IEEEauthorrefmark{2},
    Lauren Alyssa Marks\IEEEauthorrefmark{1},
    Roqaya Elmenshawy\IEEEauthorrefmark{1},\\
    Devorah Kletenik\IEEEauthorrefmark{3},
    Sean Mullen\IEEEauthorrefmark{4} and
    Rachel F. Adler\IEEEauthorrefmark{1}}
    
    \IEEEauthorblockA{\IEEEauthorrefmark{1}School of Information Sciences, University of Illinois Urbana-Champaign, USA. \{ac62, lamarks2, relme, radler\}@illinois.edu}
    \IEEEauthorblockA{\IEEEauthorrefmark{2}Windows Insider, Microsoft, Bhopal, India. bashab17@pm.me}
    \IEEEauthorblockA{\IEEEauthorrefmark{3}Department of Computer Science, Brooklyn College, CUNY, USA. kletenik@sci.brooklyn.cuny.edu}
    \IEEEauthorblockA{\IEEEauthorrefmark{4}Department of Health and Kinesiology, University of Illinois Urbana-Champaign, USA. spmullen@illinois.edu}
}
\maketitle

\begin{abstract}
We conducted a qualitative co-design study with four adults aged 60+ to gather design insights on a Figma prototype and a generative AI (GenAI) chatbot for an app aimed at providing an AI coach to support older adults' physical activity. The initial design for both incorporates several novel aspects: a curated health knowledge base, personalised responses based on goals and health history, privacy considerations, integration with wearables for physical activity context, as well as dynamic context injection. The study yielded feedback on improving both the proposed user experience in the app and the conversation flow with the chatbot, and it will aid future work aimed at implementing a GenAI-powered health coach for older adults.    
\end{abstract}


\section{Introduction}
Almost half of Americans aged 75 and older, and almost a quarter of those aged 65-75, report experiencing a disability, with challenges such as hearing, vision, mobility, and cognitive impairments~\cite{USCensusBureau2024}. Moreover, older adults, particularly those with disabilities, are more susceptible to physical inactivity~\cite{motl2010physical, harvey2015sedentary}. However, engaging in physical activity (PA) is associated with favourable outcomes on impairment, function, and health-related quality of life~\cite{motl2010physical}. 
Regular exercise among older adults can delay the onset of disabilities~\cite{wang2002postponed} and mitigate their impact~\cite{krebs1998moderate}.  

With the advancement of Large Language Models (LLMs), such as ChatGPT, there is an opportunity to harness generative AI in healthcare applications~\cite{wachter2024will, rodriguez2024leveraging}. Leveraging LLMs as conversational tools holds potential in promoting regular exercise through (1) personalised conversations tailored to their unique histories, preferences, and medical conditions~\cite{eapen2023personalization}, (2) optional voice integration for conversation, providing a more accessible means of communication, and (3)  health guidance targeted for a lay audience~\cite{zaretsky2024generative}. 

However, older adults are less likely to use eHealth services \cite{Ali}. To alleviate these concerns, it is important to involve older adults in the design process\cite{chen2025toward}. Co-design, or participatory design, is a collaborative approach in which designers and users work together throughout the design process \cite{sanders2008co} and is particularly valuable when developing health-related apps \cite{lamonica2019technology, woods2017design, adler2022developing, adler2025designing}. We co-designed a mobile health (mHealth) app together with older adults in order to ensure the resulting app would be usable, accessible, and well aligned with their needs. 

Given these considerations and the challenges that older adults face in using mHealth technologies, this paper aims to address the following research questions:

\begin{itemize}
    \item[] \textbf{RQ1}. What needs, preferences, and design features do older adults identify as valuable when co-designing an mHealth intervention for physical activity promotion?
    \item[] \textbf{RQ2}. How do older adults perceive and interact with an LLM-driven conversational interface during the co-design process, and what factors shape their acceptance or concerns?
\end{itemize}
\vspace{-5pt}
\section{Background Literature}

Co-design has become a central approach in the development of digital health interventions. It is based on the idea that end users, clinicians, and developers should work together throughout the design process to ensure that technologies align with real needs and contexts, and allows researchers to embed users’ requirements and values early on and proactively identify potential barriers to implementation~\cite{malloy2022co}. Co-design is widely used across digital health domains, although the quality and consistency of reporting remains uneven and often insufficient for assessing its true impact~\cite{kilfoy2024umbrella}, and also particularly recommended for the development of mHealth systems, due to their complex environments and multiple stakeholders~\cite{noorbergen2021using}. Indeed, co-design approaches have been used in the development of a diverse spectrum of mHealth apps, including those for patients with heart conditions~\cite{gibson2023using}, for dementia~\cite{fox2022co}, cancer~\cite{adler2022developing}, and diabetes~\cite{kwan2023mobile}. In the context of Generative AI (GenAI), there is promise for its use in healthcare but it introduces distinct risks in terms of trust, accuracy, and 'hallucinations'; these necessitate the need for rigorous, human-in-the-loop design processes~\cite{wachter2024will,rodriguez2024leveraging}.

Participatory approaches can empower older adults in particular and support the development of digital tools that better reflect their needs for independence and daily living. Systematic reviews~\cite{cole2022codesign, sumner2021co} show that older adults can contribute meaningfully as design partners, particularly when engaged in multiple co-design activities such as interviews, prototyping, and focus groups. Their participation encourages mutual learning and helps shape design decisions based on real-life experiences. Co-design with older adults also poses unique challenges: for example, older adults may have less experience with newer technology, which may limit their participation ~\cite{harrington2018designing}. Another possible challenge arises when older adult participants want functionalities that are not feasible or interface elements that contradict best design practices~\cite{tong2022lessons}. 

A scoping review by~\cite{constantin2022use} found that co-design in PA interventions takes many forms and can produce interventions aligned with older adults’ needs, preferences, and contexts. Research further shows that digital co-design is feasible and inclusive when supported by relationship-building, digital literacy assessment, tailored technical assistance, and flexible formats~\cite{darley2022conducting}. Involving participants meaningfully can encourage mutual learning, boost confidence with technology, and give them a sense of ownership over the design process~\cite{cole2022codesign}. While evidence on improvements in usability or health outcomes is still limited, co-design can help produce tools that are relevant and user-centred. One cautionary note is expressed by~\cite{blouin2025methods} that posited that involvement of older adults at all stages of the process and reflecting on the possible contribution of biases and stereotypes alone might not translate automatically into better design decisions.

\section{System Design}
Based on an earlier qualitative interview study with 20 older adults, which found initial interest in a personalised, multimodal, and interactive smartphone app targeting PA and sedentary behaviour, we proceeded to create: (1) a prototype in Figma for a PA app with a conversational agent (CA) and (2) a GenAI-based CA.
The app is intended for both informational and motivational purposes for older adults in the context of PA and sedentary behaviour. The design assumes that users would sign up with limited information about themselves, and over time, interactions with the chatbot would lead to more personalised knowledge and tailored responses. 

\vspace{-4pt}
\subsection{Prototype Design}
\vspace{-4pt}
  \begin{figure*}[t]
        \centering
        \includegraphics[width=\textwidth]{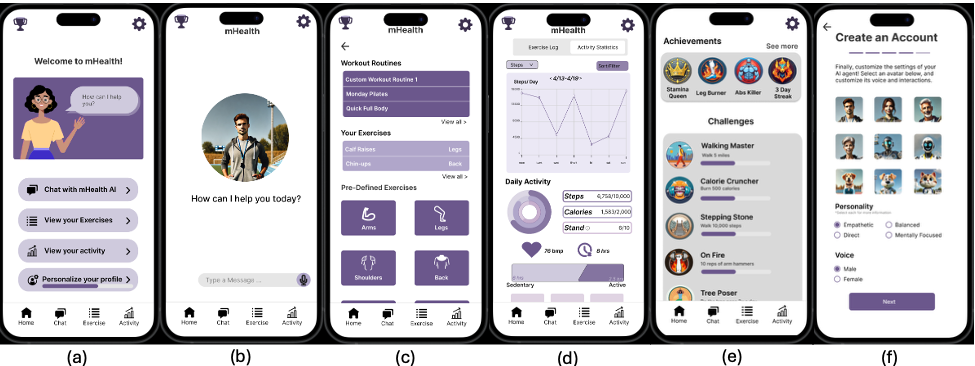}
        \caption{Screenshots of the health app, showing (a) the main screen (b) the ability to chat with a health coach, (c) view exercises,  (d) track activities, (e) gamification challenges, and (f) customise the avatar, its voice, and personality.}
        \label{app}
    \end{figure*}
    
On the main screen (Figure \ref{app}a), participants can either chat with their virtual coach (Figure \ref{app}b) or visit the Exercise tab to access the Exercise Library (Figure \ref{app}c), create a workout routine, or log an activity. The Activity tab (Figure \ref{app}d) allows users to review their fitness activities. From the trophy icon on the left corner, users can access challenges and achievements for gamification  (Figure \ref{app}e). Additionally, users can customise their virtual health coach avatar (Figure \ref{app}f). 
\vspace{-4pt}
\subsection{Conversational Agent}
\vspace{-4pt}
We wished to deploy a multimodal CA that used specialised health knowledge, privacy preserving, hallucination-resistant, and economical in both time and cost terms. The ensuing design is detailed below:

\textbf{System Architecture and Prompt Orchestration Protocol}
The mHealth chatbot operates privately on a serverless, event-driven architecture decoupling frontend interactivity from backend logic. It is both synchronous and records historical audit trail of data. Concurrently, the client triggers a server-side flow execution utilising the open-source \textit{Genkit} framework. This flow is effectively a semantic orchestrator with Retrieval-Augmented Generation (RAG). The system programmatically retrieves unstructured text data from the consolidated knowledge base and injects this context into the inference window; this process is also known as Dynamic Context Injection (DCI). This architecture constrains the generative model to produce responses grounded almost exclusively to the verified domain corpus, thereby mitigating the risk of hallucination, something critical to avoid in the health advising context.


\textbf{Knowledge base for RAG}
The knowledge base was developed through systematic curation of evidence-based exercise guidelines, safety protocols, and behaviour change frameworks relevant to older adults. Primary sources included U.S. government publications (NIH, CDC, HHS, Physical Activity Guidelines), open-access organisational guidelines (WHO, ACSM), peer-reviewed systematic reviews from open-access journals, and the investigative team's own published work. All materials were either public domain, published under Creative Commons licensing, or used with appropriate permissions. Condition-specific adaptations were drawn from consensus statements (e.g., ACSM) and validated screening tools (PAR-Q+ACSM risk stratification algorithms). Content was structured for RAG by chunking documents into conceptually discrete units (e.g., aerobic prescription parameters, resistance training contraindications, fatigue management strategies), rather than by page or section. Each chunk was tagged with metadata including target population, exercise modality, evidence level, and applicable safety considerations. The resulting corpus was iteratively tested using representative user queries, with expert review to verify retrieval accuracy and clinical appropriateness. 

\textbf{Implementing Audio Functionality for Enhanced Accessibility.}
To enhance accessibility, we implemented client-side Speech-to-Text (STT) and Text-to-Speech (TTS) capabilities. By treating voice input as an alternative serialisation format for the underlying text-based logic, we ensure that the core reasoning engine and safety protocols remain consistent regardless of the input modality.


\textbf{Dynamic Context Injection for Personalised Advising}
The system is configured to have CA responses personalised for each user but it also allows general behavioural constraints imposed by the administrator for the CA version used by any user, essentially allowing 'global' commands in place, such as "Please advise stopping exercise if user reports pain". This process involves the runtime retrieval of two distinct data schemas: the ``Chatbot Persona'' and the ``User Profile.'' The persona configuration, managed by administrators, imposes behavioural and tonal constraints (e.g., empathy, simplicity) upon the model. The user profile forms the base for informational markers to embed directly into the prompt structure, and this user profile itself is continually updated to keep current. By utilising the \textit{Handlebars} templating language to conditionally render these inputs, the system essentially `primes'the LLM's attention mechanism. This ensures that the retrieval-augmented facts are not merely recited but are syntactically and semantically adapted to the specific user's context, resulting in highly personalised and relevant health guidance.

\section{Methodology}
We co-designed with four US-based older adults (ages 60-70), who had participated in a prior interview study, in line with the UN and WHO classification of older adults as adults over 60 years old~\cite{unhcr_older_persons,who2025ageing}. All were university-educated, technology-proficient (e.g., used wearables), had varied PA levels and health history, and were familiar with CAs. They had demonstrated interest in co-designing and also in a future app developed using the process.
Co-design sessions were conducted individually and remotely using Zoom and lasted 90-120 minutes on average. These were recorded and transcribed. Our primary focus was on gleaning design insights on the user experience with the proposed app design but we also wished to test the CA flow and get feedback on the proposed personalisation using the CA as well as the integration of text and audio. Thus, the co-design sessions were divided into two parts, eliciting feedback separately on the Figma design prototype and the personalised CA. During the co-designing process, the co-designers answered open-ended questions about their daily and long-term PA practices, experiences with health apps and wearable trackers; trust, perceptions, and expectations of AI systems; reactions to prototype screens (onboarding, chat, exercise library, activity dashboard); desired features, accessibility concerns, and motivations.

The Figma design prototype displayed limited functionality, yet it enabled co-designers to visualise the proposed design. For screens using placeholder data, the research team explained the data that would be shown on screen in the actual application. The co-designers engaged in think-aloud reasoning while navigating screens.
For the interaction with the CA, prior to the study, we emailed the co-designers a list of synthetically generated user profiles and suggested the one chosen for the study. These hypothetical profiles of older adults included basic bio-data as well as health goals, health challenges, and mock past conversation history and PA. For example, the profile used was `Charlie,' who had a weekly and monthly step count as well as challenges that he was supposedly pursuing. During the co-design session, the co-designers familiarised themselves with the profile and then asked questions to the CA. They used both audio and text for input. Since the study was done remotely and the co-designers did not have direct technical access to the CA -- it was deployed locally -- the co-designers spoke the questions live and this was interpreted by the CA in real time. Alternatively, they spoke or typed the question in the Zoom chat and the research team then relayed the question to the CA locally. Thus, both STT and TTS were used.

We conducted an inductive, iterative qualitative analysis of the session notes and transcripts. Three members of the research team reviewed the materials and clustered recurring challenges, ideas, and design considerations into themes.
\vspace{-5pt}
\section{Results}

Our qualitative analysis of the co-design sessions produced nine themes that describe co-designers' feedback.

\subsection{Visual Familiarity and Predictability}
\vspace{-4pt}

We found that co-designers generally preferred having an interface similar to commercial apps or common navigation flows. They found the form completion, settings, and other navigation flows to be relatively straightforward and did not suggest changes to the overall structure. The conventional design of the Settings screen was appreciated for its predictability, despite being visually not exciting. However, for other pages, such as Achievements and Activity Statistics, the co-designers did not recommend simplifying these pages, despite the amount of information they contained, as they considered the content useful for tracking purposes. One co-designer (C3) described the Activity Statistics page as a `postmortem' of activity, and therefore expected it to include substantial detail. 
\vspace{-5pt}

\subsection{Adequate Context and Explanations}
\vspace{-4pt}
A recurring point raised by the co-designers related to the need for adequate explanations of the terms. Their primary feedback focused on the importance of providing sufficient context for the terms or actions required on each page. In the Challenges page, C2 wished that the criteria for meeting a challenge, such as needing to complete 5000 steps in a day, be made more obvious in the design. In the activity statistics page where circular rings (similar to Apple Health rings) were used to illustrate progress towards various activity goals, C4 could not make out which ring stood for which goal, and wanted this to be made clear in the design using both colour and text. Adequate context and explanation was also desired in chatbot responses, and thus link to sources were brought up during the co-designing process.
\vspace{-5pt}

\subsection{A process of iteration not `front-loading'}
\vspace{-4pt}

Co-designers preferred adding information to the app gradually over time rather than in large blocks. This issue arose in several parts of the app. For example, the initial signup form included a section for adding pre-existing health conditions. They wanted to be able to add multiple conditions separately. that they could also edit this later on. Co-designers also raised the issue of adding lengthy complicated exercise routines and the ability to save them. All the co-designers found this to be potentially cumbersome, although they said when sufficiently motivated to develop and adhere to an exercise routine, they might still decide to input a whole exercise routine in one go. They preferred and suggested the addition of flows that allowed easy saving and modification of routines. They also suggested that the motivational challenges to complete be personalised for them.

\vspace{-5pt}

\subsection{Multi-device compatibility}
\vspace{-4pt}

While conversing with the chatbot, the co-designers used audio extensively, for both input and output. However, they said that they would typically read but would use audio in real life situations when they are unable to focus on the screen, such as while cooking or driving. However, the co-designers pointed out that using text input on their mobile phone to talk to the CA would get tiring very quickly. They wished to also have the ability to converse with the CA through a computer using a keyboard because typing on a mobile phone for long conversations felt tiring. Another desired aspect was that the data from the wearable be synchronised frequently to the smartphone.      

\vspace{-5pt}
\subsection{Intelligent use of colour contrasts}
\vspace{-4pt}
The co-designers liked the use of a consistent theme throughout the app and the purple colour used was visually appealing. However, there were two notable comments: C1 suggested using some additional colours sparingly to add additional emphasis. The `Achievements' section of the app design had more colours and this was perceived to be attractive. 
However, for the very same page, C4 found the bright colours used in some icons to be overwhelming and felt that there was too much happening in the image. It became apparent that he preferred simpler icons framed against clear space in the background. Another relevant point was about the use of colours in the daily activity goal rings. The version used in the co-design study employed multiple shades of violet, but co-designers suggested that the colour differences be more distinct and less subtle.     
\vspace{-5pt}
\subsection{Use of Menus and Landing Pages}
\vspace{-4pt}

The use of a consistent bottom menu and the gear icon for settings felt clear to co-designers. However, they preferred that the  visually prominent central area of the home page not be used to redirect to parts accessible from the menu but instead serve as a navigation aid for accessing parts of the app that would otherwise need multiple taps to get to.
\vspace{-5pt}
\subsection{Multimodality}
\vspace{-4pt}
The ability to speak to the chatbot by voice and also to have the answer read aloud was considered essential when discussing design. The co-designers appreciated the implementation of these features. The position and colours of the audio button were discussed in detail by C1, \textit{"If it is to work as a click on and click off you should have it switch colors depending on the mode it is in, such as purple or gray for off and red when recording, as an indicator of its state."} A drawback of the multimodal interaction was that presentation of tables were hard to understand when read aloud, although they increased visual simplicity. Furthermore, the use of numbers in long lists sounded annoying (C4) and a user could not change how the text was read but only ask the CA to align the text to desired format and constraints through instructions; this was only partially successful in practice because the exact prompt needed for that had to be refined multiple times to get to the desired result.
\vspace{-5pt}
\subsection{Privacy affects information shared}
\vspace{-4pt}
When discussing interactions with the chatbot, co-designers raised concerns about privacy, expressing reluctance to share sensitive health data with third parties. However, they welcomed the locally hosted, secure implementation of the CA. While they were not as excited about a cloud-based version if it involved third party data sharing, they did not explicitly state that they would not use it. 

\vspace{-5pt}
\subsection{Interactive and Proactive Information Gathering}
\vspace{-4pt}

While it was expected that the chatbot would  actively learn from the user by asking questions, co-designers were less willing to manually provide information in the form of long forms for personalisation, preferring gradual data collection through interaction. Information was better gathered over time especially if the chatbot could proactively ask the user questions through notifications or queries at certain points, such as after completing an exercise when it could gauge tiredness (C3).  

\vspace{-5pt}
\section{Discussion}
Our co-design sessions with older adults revealed consistent preferences around usability, multimodal interaction, incremental personalization, and privacy. These themes align with prior work and highlight key considerations for LLM-driven mHealth tools.

\paragraph{Usability and co-design for older adults} Participants preferred predictable navigation, low cognitive load, and the ability to edit personal and health information over time instead of completing long onboarding forms. Systematic reviews show that involving older adults throughout design aligns digital health tools with users’ routines, needs, and capacities~\cite{cole2022codesign}. Our findings also support existing mHealth guidelines for older adults, including high contrast, simple layouts, minimal motor demands, clear labeling, and memory-friendly structures~\cite{liu2021mobile}, highlighting that participatory design reliably elicits age-appropriate usability requirements.

\paragraph{Audio vs. visual modalities -- the tradeoffs of multimodal output} A central insight from our sessions was the tension between designing for visual richness (e.g., detailed statistics, graphs, icons) and spoken, audio-based consumption via a CA. Voice assistants can be powerful tools for older adults, but usability challenges persist~\cite{mahmood2025voice}. What works well visually, such as detailed activity statistics pages or achievements dashboards, does not always translate to voice: long lists, numeric tables, and subtle colour differences can be hard to parse and may cause confusion or cognitive overload. Our findings align with prior work showing older adults appreciate multimodal voice + screen interfaces; for example, a real-world deployment found that although users preferred speaking, they valued visual output when available, supporting accessibility and usability~\cite{chen2023screen}.

Optimising chatbot output for visual versus auditory presentation involves tradeoffs. One approach is to provide summarised or simplified content for audio while retaining full detail on-screen. Alternatively, without affecting the informational content, the presentation style and language could adapt to the modality--audio, visual or be balanced between both--to ensure clarity and usability. For instance, a short spoken summary followed by a richer on-screen view may better suit different situational contexts (e.g., walking while listening vs. sitting and reviewing details).

\paragraph{Visual design and usability preferences}
While colour contrast is a well-established principle for older adults, co-designers valued a consistent theme with selective use of colour to highlight key elements. The default purple theme was appealing, but participants suggested adding colours sparingly for action-oriented sections like Achievements or activity statistics. Subtle shade differences, especially in goal indicators, were sometimes hard to distinguish, reflecting age-related declines in contrast sensitivity. Iconography also mattered: overly detailed or abstract icons were overwhelming, while clear, well-separated icons were preferred, consistent with prior work showing older adults interpret fewer unfamiliar icons correctly~\cite{leung2011age}.

These results highlight a challenge for generative-AI chatbots: aligning style and format with older adults’ usability preferences requires precise prompt engineering, which many users may struggle with. Since personalisation depends on dynamic context injection and user behavior (interactions, customisations, prompt use), experiences can vary widely. This raises a key design question: to what extent should designers provide sensible defaults in iconography, color, and chatbot style, rather than relying on user-driven customisation?

\paragraph{Balancing reliability and scope in health chatbots}
Grounding our conversational agent in a curated health knowledge base highlights a key tension: questions beyond the local corpus can produce "I don’t know" responses, risking user dissatisfaction. Allowing unrestricted internet-based generation increases coverage but also the risk of inaccuracies, hallucinations, or misleading advice. RAG reduces hallucinations and improves factual grounding, but evaluations show that RAG systems remain vulnerable to unsafe or decontextualised responses, especially for complex queries~\cite{baur2025development,feldman2024ragged}. Thus, designers must carefully balance coverage (topic breadth) with reliability and safety, deciding between a limited but safe knowledge base or broader yet riskier openness.

\vspace{0.5em} \noindent \textbf{Summary of Design Recommendations:} Synthesising these insights, Table \ref{tab:recommendations} outlines key design recommendations for mHealth interfaces targeting older adults, balancing usability with the affordances of GenAI.
\begin{table}[ht]
\centering
\caption{Design Recommendations for mHealth GenAI} 
\label{tab:recommendations}
\resizebox{\columnwidth}{!}{%
\begin{tabular}{@{}l p{6.5cm}@{}}
\toprule
\textbf{Focus Area} & \textbf{Design Recommendation} \\ \midrule
\textbf{Onboarding} & (1) Allow incremental data entry over time; avoid long initial forms. \\ 
                    & (2) Offer redundant methods for adding goals (voice + text). \\ \addlinespace
\textbf{Navigation} & (3) Provide web/desktop access for easier typing. \\
                    & (4) Use central landing space for high-priority navigation. \\
                    & (5) Include clear contextual guidance (popups/text) on every screen. \\ \addlinespace
\textbf{GenAI Output}& (6) Provide ``TL;DR'' summaries for audio; full detail for text. \\ 
                    & (7) Adapt phrasing for modality (spoken vs. visual). \\
                    & (8) Cite sources visibly in responses to build trust. \\ \addlinespace
\textbf{System}     & (9) Ensure frequent, seamless background synchronisation with wearables. \\ \bottomrule
\end{tabular}%
}
\end{table}
\vspace{-5pt}

\section{Conclusion}
This co-design study draws from motivated older adults seeking to customise a future application for their own personal use and thus offers pragmatic design insights for a future health app aimed at facilitating PA among older adults. These include predictable interfaces, iterative process of information sharing, adaptability in multimodality, clear contextual guidance, effective use of white space and colour contrasts, provision of multidevice usage, and simpler flows for modifying and entering custom preferences such as exercise routines or health limitations. These insights will inform the development of an mHealth app to test the viability of using LLMs to motivate and provide information to older adults.  
\vspace{-5pt}
\section{Acknowledgments}
We would like to acknowledge contributions from co-designers: Graham P Houser, Andy Kilhoffer, Anne Cook, and Roger Fredenhagen; and other contributors: Ashley Witte, Carol Gong, and David Thomas. 
\vspace{-5pt}
\bibliographystyle{IEEEtranS}
\bibliography{references}

\end{document}